\documentstyle[preprint,aps]{revtex}
\begin{document}
\draft

\title{The Quantization process for the Driven Quantum Well - \\
Perturbative Expansion and the Classical Limit}

\author{E. Eisenberg, N. Shnerb and R. Avigur}

\address{Department of Physics, Bar-Ilan University,
Ramat-Gan 52900, Israel}

\date{\today}
\maketitle
\widetext

\begin{abstract}\begin{center}
We consider the quantum mechanical behavior of a driven particle in an
infinite 1D potential well. We show that the time dependent perturbation
series is induced by the delicate non-trivial properties of the momentum
operator in this case, namely, its non-self-adjointness. Using this
expansion, we calculate the first order contribution to the cross
section and the energy gain, and discuss their classical limit. In this
limit the one-period energy gain converges to its classical analog -
the classical local
(momentum space) diffusion coefficient. Both the classical and quantum
mechanical results are compared with numerical simulations.
\end{center}\end{abstract}


\narrowtext

\def\bb{\begin{equation}}
\def\ee{\end{equation}}
\def\hf{\hbar_{\rm eff}}

\section{INTRODUCTION}
Modern semiconductor technology has enabled the fabrication of 1D
quantum wells \cite{hol1}.
Such a quantum well is fabricated by varying the
alloy composition in a compound semiconductor like $Al_x Ga_{1-x}
As$ along one dimension. Conduction electrons in such structures
experience an arbitrarily shaped effective potential in the growth
direction while remaining free in the perpendicular plane. Quantum
wells are typically $200 - 300$ meV deep with level spacing
$\Delta E$ between several meV and $150$ meV. These systems are of
special interest since they can be treated in means of pure
quantum mechanical considerations while they are still
experimentally accessible.

Recently, there has been interest in the behavior of such systems
under the influence of an electromagnetic field
\cite{mmtg,reich1,bmi,bdh,anals,rcb,rcb2,holt,fmbor}. The quantum well
structure can be considered as an analogue of an 1D atom, and thus
a study of the driven well may help us learn about the
interaction of atoms and high-field electromagnetic radiation. In
the region where the electric field is strong with respect to
the level spacing of the well, one obtains a system in which to
study non-perturbative effects in light-matter interaction.
Quantum wells with parameters in the region of interest in this
study are well in the range of experimental capabilities. In fact,
currently there are some experiments carried out at UCSB with
the Free Electron Laser (FEL) in which an intense monochromatic
far-infrared radiation is applied to a quantum well \cite{rcb2}.
In particular, some authors have studied in the recent years the
quantum dynamics of a particle in an infinite potential well
driven by a spatially constant ac electric field \cite{rcb,rcb2}.
These studies
analyzed the energy absorption characteristics as a function
of the driving force amplitude and frequency. A non-perturbative
resonance was found for a special value of the driving
amplitude, corresponding to the replacement of the states localized
in phase space by more extended states. This resonance was
predicted by avoided crossings in the quasienergy spectrum of the
Floquet operator. Other mesoscopic applications of this model have been
also considered \cite{fmbor}.

The problem of a driven particle in an infinite potential well
is of interest in the field of quantum 'chaology'
\cite{ber1} as well.
The study of the effect of quantization on the properties
of classically chaotic systems has attracted increasing
attention in the recent years \cite{book,hakke,reich,kenes}.
It was argued long ago that quantum chaos is strictly impossible in
autonomous finite quantum systems which exhibits a discrete energy
spectrum \cite{pen}. Moreover, a peculiar effects of quantum
suppression of classical chaotic diffusion in the dynamics of
the kicked \cite{cas1,fish} and two-sided kicked \cite{us1,us2}
rotator has been shown to exist.
Since our system is classically chaotic, the study of its quantum
properties may enrich the search for the fingerprints of classical
chaos in the quantized system. A WKB approach was employed to this
system, and information about the quasienergy levels and the
quasistates is available in the classical limit.

In this paper we consider the quantum mechanical behavior of the
above model, i.e., a driven particle in an infinite potential well.
We discuss the proper way to apply a perturbative expansion
to the system, and explain in detail the origin of the failure
of the more naive approach. Our treatment applies also
strong fields in the high frequency limit. We now turn to a
concise qualitative description of our approach.

Applying a guage transformation, one can show that the
Hamiltonian can be written as a function of the operators $P$
and $P^2$ alone with time-dependent coefficients.
Thus, it seems that the Hamiltonian commutes with itself for
different times, and one may simply integrate over the time
to get the evolution operator. The resulting evolution operator
for a complete period (and a number of periods as well)
behaves as one which follows from an integrable Hamiltonian.
Moreover, for a properly chosen period, i.e., a proper choice of
the start (and end) point, the complete period behavior is
exactly the same as that of the unperturbed well.

This result is physically unacceptable. It is well known
\cite{z78,izr}
that the quantum dynamics of a system corresponds to its
classical dynamics for time periods which scale (at least)
like $\ln(1/\hbar)$. This result can be rigorously proved
using the path-integral formulation of the expansion
around the WKB limit. Accordingly, since the classical
driven particle experiences a chaotic motion,
one should expect the quantum dynamics to show some features
that resembles this chaotic (diffusive) behavior in the
short time limit (although the long time limit is charaterized
by the energy-space localization which is a purely quantum
mechanical effect). Nevertheless, if the quantum dynamics
is determined by the integrated Hamiltonian,
which is exactly equal to the one obtained from an integrable
system, namely, the unperturbed well,
it must also describe in this short time limit
the classical dynamics related to this integrable Hamiltonian.
However, the classical dynamics of these two are off course
qualitatively differenet; for example, while the first diffuses
over the whole stochastic layer, the second is energy-preserving.
Note that the above described derivation is exact for any
value of $\hf$ (the effective $\hbar$). Thus one may choose
$\hbar$ to be arbitrarily small, such that the correspondence time
diverges. One may therefore conclude that the result of this
derivation contradicts the well-established 'log-time'
correspondence principle.

It turns out that the origin for this apparent contradiction
is found in the delicate non-trivial properties of the
$P$ operator, when defined on a finite interval. We discuss
some of the manifestations of this behavior, and explain
in detail its implication on our derivation. It is shown
that the seemingly trivial relation $[P,P^2]=0$ does not hold
for a particle in an infinite potentail well, and therefore
our guaged Hamiltonian {\it does not} commute with itself at
different times. An explanation to this non-trivial behavior
is given in terms of the self-adjoint extension procedure
relevant to our model. This unexpected non-commutativity
may be of importance to the analysis of many other
finite quantum systems. In particular, each finite system
under the influence of an electric or magnetic fields
may be influenced by this property.

In order to study the effect of this non-commutativity
of $P$ and $P^2$ over the dynamics, we use the explicit form
of the operator $[P,P^2]$ in the energy basis, and employ the
time-dependent perturbation theory to the interaction
representation of the guaged Hamiltonian, which takes into
account an infinite number of commutators
$[P^2,[P^2,[\cdots,[P^2,P]\cdots ]]]$, and obtain the first
correction to the exact periodicity implied by the previous
false derivation. Our expansion is appropriate even for cases
in which $E_L=\lambda L$, the (maximal) potential difference
between the ends of the well, is much larger then $E_0$, the
spacing of the energy levels in the unperturbed well. The matrix
elements of the perturbation are not small and in fact, the
reason that the usual time-dependent perturbation theory does
not break down is the approximate periodicity
implied by the preceding wrong argument. We thus obtain an
expression for the integral cross section, i.e., the probability
to leave the initial state after one period of the perturbation.
The high frequency limit of the frequency dependence of this
cross section is $O(\omega^{-4})$, i.e. much faster then the usual
$O(\omega^{-2})$ asymptotics, which characterizes the response in
the high frequency limit.

The first-order treatment yields an infinite series for
the escape probability - the probability {\it not} to
return to the initial state after a complete period, and
the energy gain. We discuss the asymptotics of these
serieses in the various regimes. In particular, we study
the behavior in the classical limit, i.e., $\hf\to 0$ while
keeping the energy fixed. We show how the approximate periodicity
breaks down in this limit, thus furnishing the transition to
the classically chaotic behavior. The classical limit of
the energy gain is then compared to the corresponding classical
quantity. We find that, surprisingly, the indirect calculation
of this quantity, i.e., evaluating the quantum result and then
considering the classical limit, is much more efficient then
a direct calculation based on the classical equations of motion.
The first order results also imply the exsistence of an
antiresonance phenomenon, namely, a sharp anti-peak in the
escape probability and the energy gain for some values
of $\hf$. This phenomenon was discussed in detail recently
\cite{wellres},
and it was shown that it persists for higher orders of
perturbation theory as well. This was tested numerically,
and it was found that the antiresonance behavior is seen
even for high values of $\beta$ the perturbation strength
parameter, even up to $\beta=2$. Since the full discussion
was given recently elsewhere, we present here only a
short descriptive consideration of this effect.

The rest of this paper is organized as follows: In section II we
present the naive 'exact' solution of the model in terms of the
transverse guage. The surprising result is that the effects of
the ac field are cancelled out over a complete
period. We discuss the implications of this result.
In section III we explain the origin of the error in this
derivation in terms of the algebraic properties of the
operators involved with respect to their self-adjoint extension.
Section IV presents a systematic perturbative expansion
around the limit in which the naive argument holds. This
derivation is based on the interaction representation of the
$P$ operator. We apply time-dependent perturbative methods
to calculate the first correction to the
apparent exact periodicity. Asymptotic analysis of these
perturbative rsults are discussed, with particular emphasis
on the classical limit. Numerical results are presented,
which verify the above perturbative derivation and test its
range of applicability. Section V describes in short the
anti-resonance phenomenon, and a conclusion is given in section VI.

\section{FIRST CONSIDERATIONS}

The Hamiltonian considered is
\begin{equation}
H = \frac{{\bar P}^2}{2m} + eEf(\omega t)\bar X,
\end{equation}
where $f(t)$ is a periodic function whose period is
$T=2\pi/\omega$. The wave-function is imposed to satisfy
the boundary conditions of vanishing of the wave-function
at the ends of the well $\bar x=0;\ \bar x=L$.
This model has been investigated classically for the case
$f(t)=\cos(\omega t)$ \cite{mmtg,reich1,bdh}, and has been found to be
chaotic. Using dimensionless form, i.e.,
\bb
X = \bar X/L;\quad \tau = \omega t;\
\ee
it can be shown that the classical behavior of
this model depends on a single parameter $\beta=eE/(m\omega^2L)$.
However, the quantum evolution depends also on an additional
parameter, the effective $\hbar$, $\hf=\hbar/(m\omega L^2)$.
Using the dimensionless form the Hamiltonian reads out to be
\bb
H = \frac{P^2}{2} + \beta f(\tau) X,
\ee
where the position representation of the scaled momentum
operator is $P = -i\hbar_{eff}\frac{d}{dx}$.
We now transform to the transverse gauge, in which the Hamiltonian
takes the form
\bb
H= \frac{1}{2}(P - \beta F(\tau))^2
\label{hamil2}
\ee
with
\bb
F(\tau)=\int_0^\tau dt f(t).
\ee
The lower limit of the integral is $t=0$ since we choose the
field to be turned on at $t=0$, and thus the vector
potential should vanish at $t=0$.
At this point, one may naively use the following argument: since
the hamoltoniam (\ref{hamil2}) commutes with itself at different
times, the Schr\"odinger equation may be trivially integrated
over $\tau$, yielding the evolution operator $U(0,\tau =
\exp(-i H_{eff}(\tau))$ where
\bb
\label{heff}
H_{eff} = \int_0^\tau H(t) \  dt =  \frac{P^2 \tau}{2}
- \beta P \int_0^\tau F(t)  \ dt + {\rm  const}
\ee
The diagonalization problem for this effective Hamiltonian is
rather trivial since for each $\tau$ it is just the Hamiltonian of
a particle in an infinite well with a guage term, whose
eigenfunctions are of the form $\sin(kx)\exp(i\gamma(\tau) x)$.

A basic tool for analyzing the chaotic features of classical
systems is the Poincare section \cite{book,reich}.
For a time-periodic Hamiltonian system with
one degree of freedom, the Poincare surface of section is just a
strobe plot, that is one plots $(x,p)$ once every period $\tau$.
Such a plot for the driven well system is given in Figure
particular, if the evolution of a system is  periodic, the strobe
plot will give a one-dimensional curve. On the other hand if the
evolution is chaotic, the trajectory will spread over a
two-dimensional region. Accordingly, most of the research interest
in the field of quantum chaos was directed to the study of the
Floquet operator, which determines the quantum time evolution of
periodic systems for times which are integer in the driving force
period units.

Apparently, one sees that for our model, namely, the driven
one-dimensional potential well, the explicit form of the Floquet
operator can be easily obtained from Eq. (\ref{heff}) to be
\begin{equation}
\label{floq}
F=H_{\rm eff}(\tau=2\pi) = P^2 \pi +
\beta P \int_0^{2\pi}dt F(t) + {\rm const}.
\end{equation}
One ontains that the stoboscopic behavior of the system is rather
trivial, i.e., equivalent to that of a particle in an infinite
well subjected to a constant electric field. Moreover for an
appropriate choice of the function $f(\omega t)$, e.g.
$f(\tau)=\cos(\tau)$ the guage term vanishes, and the effective
Hamiltonian has no intraction dependence.

The above result is completely unacceptable physically, since it
contradicts the correspondence principle for arbitrary short times
(with respect to $\hbar$). In particular, although the classical
particle in the well reaches (in general) a completely different
point in phase space after one period of the perturbation, we have
just shown that the quantum particle develops into the state it
should have been developed in the absence of the perturbation.
Thus, according to the correspondence principle, for $\hbar\to 0$
the quantum mechanical state should correspond to {\it two
different} classical states, i.e., those points in phase space to
which the system arrives in the presence and in the absence of the
driving force. This ambiguity manifests the apparent breakdown of
the correspondence principle (at least) for times of order of the
period, no matter how small is $\hbar$. We therefore turn now to
discuss the origin of the failure of our derivation.

\section{ALGEBRAIC PROPERTIES OF THE $P$ OPERATOR}

In order to find the problematic point in the above arguments, we
state explicitly the assumptions made above. Apart from simple
algebra, we used two basic relations, i.e.,
\bb
[X,P]=i\hf;\quad\quad [P,P^2]=0.
\ee
The first assumption was used to perform the guage transformation,
and the second in the statement that the guaged Hamiltonian
(\ref{hamil2}) commutes with itself for different $\tau$s.

It seems that these two relations are well established. The first
is the basic principle of quantum mechanics, whose failure leads
to the breakdown of the whole method of canonical quantization in
the system. The second assumption is just a manifestation of the
associativity of operators multiplication. However, this second
assumption turns out to be false as we now explain in detail.

The simplest way to realize the failure of the relation $[P,P^2]=0$
is through the matrix representation of these operators in the
energy basis of the unperturbed well. The matrix elements of these
matrices are given by
\bb
(P^2)_{mn}=\hf^2\pi^2 m^2 \delta_{mn}
\ee
\bb
P_{mn}=\left\{
\begin{array}{lr} -i\hf\frac{4mn}{m^2-n^2} & {\rm m+n\ odd}\\
			0                       &{\rm m+n\ even}
\end{array}\right.
\ee
One clearly sees that these matrices do not commute. In fact the
commutator gives
\bb
([P^2,P])_{mn}=-4i\hf^3\pi^2 mn.
\ee
In terms of this matrix representation,
the reason for the un-associativity of the matrix multiplication
is that this multiplication involves summations which do not
absolutely converge, and thus the sum depends on the order of
summation. However, since one expects operators to be associative,
the question still holds whether this strange result really
manifests an operator property (and if so - why ?), or is only an
artifact of the particular basis used.

First, we note that the $P$ operator
possesses some delicate properties when
restricted to a finite interval. Physically, due to the
uncertainty principleand the finite size of the well, an exact
measurement of the momentum is impossible, and thus there is no
eigenstate of $P$ in the Hilbert space as can be seen also by
applying the appropriate boundary conditions to the differential
eigenvalue problem. Moreover, it can be easily seen that since $P$
is canonically conjugate to $X$, it generates spatial
translations.
Thus its domain is restricted to functions which do not leave the
well for infinitesimal translations.
A well known criterion for the commutativity of two operators
is the exsistence of a basis in which the two of them are diagonal.
As we have seen, $P$ has an empty spectrum and thus cannot be
diagonalized in any basis. In particular it is not diagonal in the
basis in which $P^2$ is. Thus they do not commute.

The reason for this un-associativity is as follows. The operator
$P^2$ in the Hamiltonian is not really defined as $P\cdot P$ but
rather as the self-adjoint extension of this form. $P$ itself is not
essentially-self-adjoint operator when restricted to a finite
interval \cite{simon}, and has therefore no self-adjoint
extension \cite{rem}.
Thus, although the ''bare'' differential operators $P$ and $P^2$
(i.e., the operators defined only the (twice) differentiable
functions whose derivative vanishes on the boundaries, without the
self-adjoint extension) commute on their common dense domain,
after the procedure of self-adjoint extension, the domain of $P^2$
is much larger then that of $P$. If there had been such an
extension for the $P$ operator, its square would have been
the extension of the $P^2$ operator (uniqueness of the extension).
However, since due to the above (both mathematical and physical)
reasons, $P$ can not be extended to a self-adjoint form,
the ''bare'' $P$ does not commute with the extended $P^2$.

In the next section we study the perturbative expansion with
respect to the terms obtained due to the uncommutativity.

\section{PERTURBATIVE APPROACH}

The starting point for the perurbative expansion is the
dimensionless form of the guaged Hamiltonian
\bb
H=\frac{(\hf P-\beta F(\tau))^2}{2}=\hf^2\frac{P^2}{2}
-2\hf\beta F(\tau)P +{\rm Const},
\ee
where $P$ here is the dimensionless momentum operator, whose
matix elements in the energy representation are
\bb
(P^2)_{mn}=  \hf^2 m^2\pi^2 \delta_{mn}
\ee
\bb
P_{mn}=\left\{
\begin{array}{lr} -i\hf\frac{4mn}{m^2-n^2} & {\rm m+n\ odd}\\
			0                       &{\rm m+n\ even}
\end{array}\right.
\ee

We now take into account the fact that $P$ and $P^2$ do not
commute, and treat the $P$ term as a time-dependent perturbation
to the Hamiltonian $H_0=\frac{P^2}{2}$. We use the
interaction picture in which
\begin{eqnarray}
P^{I}_{mn}(\tau)&=&\biggl(\exp^{i\tau H_0/\hf}P
\exp{-i\tau H_0/\hf}\biggr)_{mn}\nonumber\\
&=&\left\{
\begin{array}{lr}
-i\hf\frac{4mn}{m^2-n^2}\exp(i\alpha\tau (m^2-n^2)) & {\rm m+n\ odd}\\
			0                       &{\rm m+n\ even}
\end{array}\right.,
\end{eqnarray}
where $\alpha=\hf\pi^2/2$. The latter equation can be derived
simply by taking the matrix element of the whole expression
in the middle of the above equation, and
applying the exponential operators to the bra and ket states (These
are eigenstates of the exponential operators). However, another way
to look at this equation is as a sum over the commutator series
as follows
\begin{eqnarray}
\biggl(\exp(i\tau H_0/\hf)P\exp(-i\tau H_0/\hf)\biggr)_{mn} \nonumber \\
= \sum_{n=0}^\infty\frac{(i\tau/2)^n}{n!}
[P^2,P^2,\ldots[P^2,P]\ldots]_{mn} \nonumber \\
= \sum_{n=0}^\infty\frac{1}{n!} \frac{mn}{m^2-n^2} (i\tau\alpha)^n
(m^2-n^2)^n\nonumber\\
=\frac{mn}{m^2-n^2}exp(i\alpha\tau(m^2-n^2))
\end{eqnarray}
It is therfore clear  that in the perturbative approach one contain
an infinite number of commutators in the derivation, and thus
the problems described in the previous sections are avoided.

We demonstrate the approach for the calculation of the
escape probability, i.e. the probability to leave
the initial state $m$ after one period of the perturbation,
to second order in $\beta$. The amplitude for a transition
to the $n$ state is given by
\bb
\label{1order}
A(m\to n)=\beta P_{mn}\int_0^{2\pi}d\tau \sin(\tau)\exp(i\alpha\tau
(m^2-n^2)).
\ee
The escape probability (integral cross section) is therefore given by
\bb
\sigma=\sum_n |A(m\to n)|^2 = 32\beta^2m^2\sum_{\rm m+n\ odd}\frac{n^2}
{(m^2-n^2)^2} \frac{1-\cos(2\pi\alpha(m^2-n^2))}
{(\alpha^2(m^2-n^2)^2-1)^2}.
\ee
Figure ~\ref{exact} presents a comparison of this first order result
for the escape probability and the energy gain after one period,
with a full numerical solution of Schroedinger equation obtained
through a quality control Runge-Kutta method. The
agreement in both cases is quite good.

In the following, we estimate this sum for the various
relevant regimes.

\subsection{Low-Energy Regime}
We first consider the regime in which the sum is appropriately
estimated by taking the continuous limit (this regime is
going to be defined precisely later), i.e., replacing the
sum by the integral
\bb
\label{intform}
\frac{\sqrt{\alpha}}{2} I = \frac{\sqrt{\alpha}}{2}
\int_0^\infty f(x)dx =
\frac{\sqrt{\alpha}}{2}\int_0^\infty \frac{x^2}{(z-x^2)^2}
\frac{1-\cos(2\pi(z-x^2))}{((x^2-z)^2-1)^2} dx
\ee
where $z=\alpha m^2$, and $x$ was defined through
$x=\sqrt{\alpha}n$. We now consider two different subregimes.
For $z <<1$, one may estimate the integral by its value for
$z=0$, which is $I=4.663...$. The
condition for replacing the sum by the integral is that
the spacing of the summation points, i.e., $2\sqrt{\alpha}$
is sufficiently small compared to the scale of changes in
the function $f(x)$. In this subregime this condition
is equivalent to $\sqrt{\alpha}<< 1$. This, together with
the condition $z << 1$ implies $E=2z\alpha/\pi^2 <<1$.
Thus the regime in which the integral approach is
appropriate is the low energy regime. However, where
$z<<1$, the condition $\sqrt{\alpha}=\sqrt{z}/m <<1$
holds trivially since $m$, the quantum number is a
positive integer $m\geq 1$. Thus the limit $z<<1$
is relevant only to the low-energy regime.

For the $z>>1$, one should notice
that the integrand in (\ref{intform}) is a highly peaked
function around the region $x^2=z\pm 1$ (See figure ~\ref{fx}.
The maximum of the function is obtained at $x^2=z$
and its value is $2\pi^2 z$,
while for $x^2=z\pm 1$ the integrand is $\frac{\pi^2}{4}(z\pm 1)$,
i.e., about a one fourth of the maximum. The decrease of the
function away from this region is $O((x^2-z)^{-6})$, and
thus one may neglect the contributions to the integral from
any region other than the maximum. The width of the peak is
therefore approximated by the distance between the two points
$x^2=z\pm 1$, i.e. $\Delta=1/\sqrt{z}$. Thus one obtains
for the $z>>1$ regime the estimate $I\sim 2\pi^2\sqrt{z}$.
Numerical integration shows that while the $z$ dependence
of the integral is as stated above, a more accurate
value of the coefficient is
\bb
I = 14.8044\sqrt{z}.
\ee
The condition for the validity of the integral approximation
in this case is that the spacing of the sum is less then
the peak width, i.e. $2\sqrt{\alpha}<< 1/\sqrt{z}$
which implies $E=2z\alpha/\pi^2 << 1$.

Accordingly, the following results for the escape probabilities
in the low energy limit are obtained
\bb
\label{s-low}
\sigma=\left\{\begin{array}{lr}
6.8058\beta^2 E/\hf^{3/2}   & z <<1 \\
9.7268\beta^2 E^{3/2}/\hf^2 & z >>1
\end{array}\right.
\ee
In figure ~\ref{sigma-low} we show the results of a
numerical summation of the 1st order series, compared to
the two asymptotes. The transition between the two regimes
is clearly seen.

The nature of the transition from the extreme quantum regime
where the system is almost periodic (as implied by the wrong
argument of Sec. II) to the classical regime where the dynamics
turns out to be chaotic over larger and larger time scales can
now be clearly seen. As one approaches the classical limit
$\hf\to 0$ the approximate periodicity of the system is lost,
since the escape
probability, i.e. the deviation from periodic behavior,
diverges as $\hf$ approaches zero. One may also define the time
scale after which the periodicity is ruined as the inverse
of the escape probability, i.e.
\bb
\tau_{chaotic} \sim T/\sigma
\ee
and this time scale vanishes in the classical limit.
It is interesting to note that while the exsistence of an upper
limit (in time) to the correspondence principle to hold
is well established and discussed in the recent years
\cite{z78,izr}, the system under consideration
exhibits the opposite type of behavior, i.e., a lower
bound on the time in which correspondence holds.

We now apply similar arguments to the calculation of
the energy change of the driven particle after one
period. This quantity has a concrete classical meaning
and thus this example may further clarify the nature
of the classical limit. Moreover, the energy absorption
of the quantum well is also an experimentally
measurable quantity, and thus one may study experimentally
the complete transition from the extreme quantum regime
to the classical limit, by changing the parameters of
the well such that $\hf$ is in the appropriate regime.
The arguments concerning the integral estimations are
quite similar to those described above, and the results are
\bb
\Delta E=\left\{\begin{array}{lr}
20.426\beta^2 E/\sqrt{\hf} & z <<1 \\
17.772\beta^2 \sqrt{E}   & z >>1
\end{array}\right.
\ee
Figure ~\ref{energy-low} shows the energy gain per
period, as a function of $\hf$, as obtained from
a numerical summation of the appropriate 1st order
sum. The two asymptotes are also indicated on the
graph, and the agreement seems quite good.

One can clearly see that for a fixed energy $E$,
as $\hf\to 0$, $z= E/\hf $ approaches infinity.
Thus the relevant regime to the classical limit is the
second one, in which the quantity converges to the
well-defined limit $\Delta E\ ({\rm classical})
= 17.772\beta^2\sqrt{E}$. It is interesting to note
that we have thus obtained a {\it classical} prediction
based on {\it quantum} calculations. This has to be
appriciated considering the fact that the
''quantum'' calculation is tractable analytically,
as we have just demonstrated. The functional form
can be completely deduced theoretically, and even
the accurate value of the coefficient is obtained
through only one numerical integration. A direct
''classical'' calculation, i.e., one which is based
on classical equations of motion is by far more
complicated. As far as we know, the best way to
calculate the local momentum-diffusion constant
is using a Monte-Carlo method, which involved
averaging over a large number of paths related
to different initial conditions. In figure
Monte-Carlo calculation for the local diffusion
constant as a fuction of $\beta$. The functional
form is indeed $O(\beta^2)$ as implied by the quantum
calculation and the value of the coefficient
(scaled by $\sqrt{E}$) is $17.77\pm .04$. Note that for
this degree of accuracy about $10^8$ paths were
considered for each case, and the whole calculation
consumed many CPU hours. We therefore conclude
that we have shown in this example not only that
the classical limit is obtained correctly from the
quantum calculations, but also that the latter
may turn out to be by far more efficient.


We wish to note here about one more point. Considering
the high frequency limit of this problem, it is clearly
seen that while $\omega >>1$, both $\hf$ and $\beta$
approach zero. Thus, the pertubative, classical-limit
calculation holds and one may conclude that the
high-frequncy asymptote of the absorption is $O(\omega^{-4})$
rather than the usual $O(\omega^{-2})$.
This is another manifestation of the approximate periodicity
of the system in the extreme quantum regime.

\subsection{High-Energy Regime}

As we have shown above, in the high-energy regime
only the limit $z>>1$ should be considered. We can then
maintain the picture of a highly peaked function $f(x)$
as in the low-energy case. However, here the function values
are taken in large steps compared to the peak width. Since
the decay of $f(x)$ away from the peak is extremely fast,
we can approximate the sum by the contribution of the
closest points, i.e., $n=m\pm 1$. This implies
(for simplicity we assume that the initial state
$m$ is not close to the ground state, $m>>1$)
\bb
\label{high-sig}
\sigma =
\frac{32\beta^2 E}{\hf^2\pi^2 (2\pi^2E-1)^2}(1-\cos(\pi^2\sqrt{8E})).
\ee
The $E$ dependence of $\sigma$ ($\hf$ fixed)
is presented in figure ~\ref{Edep}, as obtained from summation of
the first order series.
The transintion between the three different regimes is clearly
seen. For low energies, $z<<1$ and the behavior is $O(E)$.
In the intermidiate regime $E$ is still small, but $z$ is large
and one obtains a $O(E^{3/2})$ dependence
(See Eq. (\ref{s-low})). For $E>1$ the oscillatory behavior
of Eq. (\ref{high-sig}) dominates. The peaks height decays
like $E^{-1}$, as implied by Eq. (\ref{high-sig}). The straight
dotted line corresponds to the function
$$\frac{64\beta^2 E}{\hf^2\pi^2 (2\pi^2E-1)^2}$$
and passes through the maxima obtained from Eq. (\ref{high-sig}).

A similar analysis is done for the energy absorption.
Again, we calculate the classical limit of the
quantum result and get
\bb
\Delta E = \frac{16E(2\sqrt{2E}\pi^2\sin{2\pi^2\sqrt{2E}}(2\pi^2E-1)
-(2\pi^2E+3)(1-\cos{2\pi^2\sqrt{2E}}))}{(2\pi^2E-1)^3}.
\ee
This result is obtained only through the above analytical argument
and much tedious but straight forward series expansions.
This is to be compared to the standard, ''direct'', classical approach
which calls for massive computer simulations.

\section{ANTI-RESONANCE}

The antiresonance phenomena in the quantum driven well
have been discussed elsewhere. We therefore just
scetch the general arguments and main results.

The first order calculation indicated the possibility
of an antiresonance to occur. Under the antiresonance
conditions, a sharp anti-peak in the escape probability
and the energy absorption is expected. The antiresonance
is obtained when the nominator in Eq. (\ref{high-sig})
vanishes (See also fig. ~\ref{Edep}).
It can be easily checked that this corresponds
to the fact that the particle passes the well an integer
number of times during one period of the perturbation.
Recently, it was shown that this anti-resonance is
maintained even in higher orders of perturbation theory
and numerical integration of Scroedinger equation
was carried out and showed that it is stable for
$\beta$ values as high as $\beta=2$.

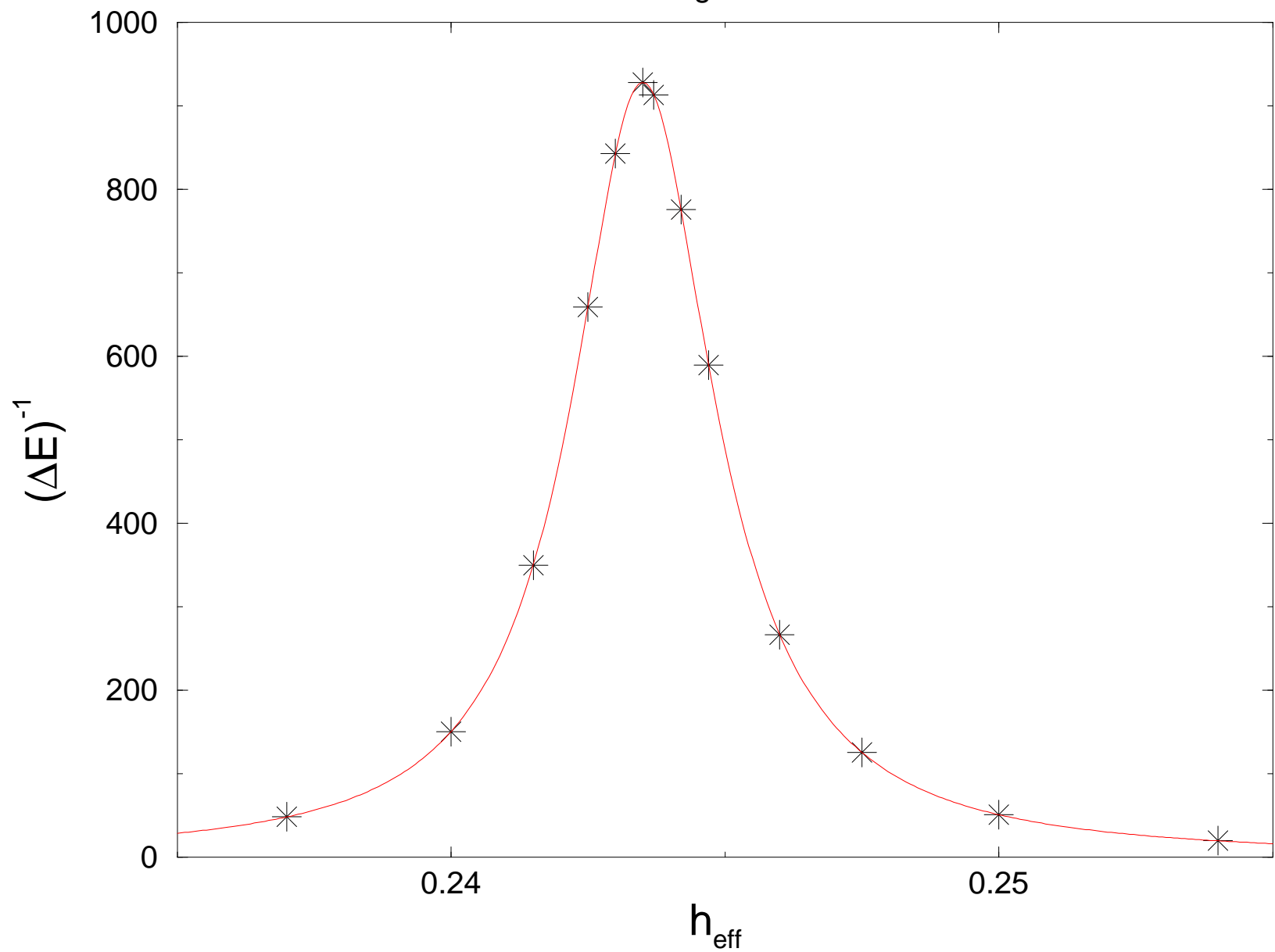
\begin{figure}
\caption{The inverse of the energy absorption per period vs. the
effective $\hbar$, as obtained from full numerical integration of the
Schrodinger equation. The stars are the numerical results, and the line
is used
just to lead the eye. The initial state is the ground state,
and $\beta=2.0$. The anti-peak in the energy absorption correspondes
	to the anti-resonance.}
\label{antir}
\end{figure}

In order to complete the picture, we present here similar
results for the energy absorption around the resonance.
These are shown in figure ~\ref{antir}.
The numerical integration of Schroedinger equation was
done using a quality-controlled Runge-Kutta method.
This sharp peak (even for large $\beta$) should be easily
detected experimentally.

\section{CONCLUSION}

The ac-driven particle in an infinite potential well is
considered. This system, which is classically chaotic
presents a rich quantum behavior.

In the first sections we have
shown that all the non-trivial properties of the system are
related to the fact that in the infinite potential well
the relation $[P,P^2]=0$ does not hold. If it did, the
evolution would become exactly periodic, which contradicts
the chaotic behavior in the classical limit. It is
really worth noting that here the {\it classical}
limit depends upon some {\it non-commutativity} property,
which seems to be a pure quantum-mechanical effect.

Apart from being crucial for the analysis of our system,
the non-commutativity of $P$ and $P^2$ may turn out to
be of importance in many cases in which the response
of quantum wells and other finite quantum systems
to electric and magnetic fields. For example, it is
interesting to check the influence of this uncommutativity
on the Landau quantization of particles in a finite box.
This problem is closely related to the investigation of
the quantum Hall effect with realistic, i.e., non-periodic,
boundary conditions.

In the second part of the paper, a perturbative expansion
around the limit in which the system is periodic was
developed. The various regimes were considered, and
the analytical estimates for the different asymptotes
were obtained. These were tested with respect to an
exact summation. The classical limit was also discussed.
It was shown how the approximate periodicity breaks down
as $\hf\to 0$. A classical limit for the absorption
was obtained from the classical limit of the quantum calculations,
and it was shown that it agrees with the classical
result obtained from integrating the classical
equations of motion. Moreover, the indirect calculation
of this quantity, i.e., taking the classical limit of
the quantum result, turns out to be much more efficient
then the ''pure'' classical one. In fact, while the first
is done analytically, and involves only one numerical
integration for determining a numerical factor, the latter
is done through a time-consuming and not very efficient
Monte Carlo simulation. Another feature of this quantum
system is the antiresonance phenomena for which the
system evolution turns out to be almost periodic for
some sharp values of the parameter $\hf$.

As stressed in the Introduction, this system, including
all the above various regimes, is experimentally
realizable, using quantum wells radiated by a laser.
Different values of the width of the well and the laser
frequency controll the parameters $\hf$ and $\beta$
such that the whole parameter space is accessible.
The influence of perurbations caused by impurities and
imperfections in the well, dissipation due to phonon excitation,
finiteness or distortion of the well, motion in the
plane of the well, and even electron-electron interactions
were discussed in Ref. \cite{rcb2} and further publishings,
and it is claimed that the one-particle, 1D, ideal system
effects can be in fact realized.
It is therefore our belief that this system may
become the subject of intense experimental, as well as
theoretical, study, as a prototype of driven,
classically-chaotic mesoscopic systems.

\acknowledgements{We are grateful to J. Avron, A. Baram, I. Dana,
L.P. Horwitz, Y. Kannai and M. Rosenbluh for most valuable discussions
on various aspects of the model.}

\begin{figure}
\caption{Stroboscopic map of phase space trajectories for the linearly
driven well, for $\beta=0.01$, and several initial conditions.}
\label{poincare}
\end{figure}

\begin{figure}
\caption{A Comparison of the numerical summation of the
	first order expression (line), with the numerical integration
	of Schrodinger equation (circles). $\hf=0.02$ and $\beta=0.005$.
	(a) The escape probability. (b) The energy gain.}
\label{exact}
\end{figure}

\begin{figure}
\caption{The function $f(x)$, the integrand
used for the integral approximation, for various values of $z$.
The function is scaled by a factor $1/z$}
\label{fx}
\end{figure}

\begin{figure}
\caption{Exact summation of the first order expression for
	the escape probability, compared to the two asymptotes.
	The stars correspond to the exact summation results
	and the lines to the asymptotes. The (dimensionless)
	energy is $5\cdot 10^{-5}$}
\label{sigma-low}
\end{figure}

\begin{figure}
\caption{Same as the previous figure for the enrgy gain.
	The thick line correspond to exact summation results
	and the straight thin lines to the asymptotes.}
\label{energy-low}

\end{figure}

\begin{figure}
\caption{The energy gain after one period, as obtained from
	averaging over $8\cdot 10^7$ classical paths at energy
	$E=5\cdot 10^{-5}$ (stars). The line is the best fitted
	$\Delta E = A\beta^2$ line, and the coefficiet $A$ is
	$17.77\pm0.04$.}
\label{clas}
\end{figure}

\begin{figure}
\caption{The escape probability $\sigma/\beta^2$ as a function of $E$.
The effetive $\hbar$ value is $0.02$. The right dotted line corresponds
	to the maxima values in the high-energy regime, as obtained
through the closest points approximation. The other lines correpond
	to the two low-energy asymptotes.}
\label{Edep}
\end{figure}

\end{document}